\documentclass[reprint,superscriptaddress, amsmath,amssymb, aps, prresearch, longbibliography, floatfix,nofootinbib]{revtex4-2}
\usepackage{amsmath}
\usepackage{graphicx}
\usepackage[normalem]{ulem}
\usepackage{bm}
\usepackage{natbib}
\usepackage{dsfont}
\usepackage{tikz}
\usepackage{epsfig}
\usepackage{feynmf}
\usepackage{blindtext, rotating}
\usepackage{mathtools}
\usepackage{dsfont}
\usepackage{subcaption}
\usepackage{physics}
\usepackage{amsfonts}
\usepackage{xcolor}
\usepackage{ragged2e}
\usepackage{siunitx}
\usepackage{comment}
\usepackage{soul}
\usepackage{empheq}
\usepackage{lipsum}
\usepackage{hyperref} 
\hypersetup{breaklinks=true, colorlinks=true, citecolor=blue, linkcolor=cyan, urlcolor=blue,filecolor=blue}
\usepackage{orcidlink}

\usepackage[T1]{fontenc}

\usepackage{xcolor} 

\DeclareCaptionJustification{justified}{\justifying}

\DeclareSIUnit{\rad}{rad}

\definecolor{bright_blue}{HTML}{85C1E9}
\definecolor{middle_blue}{HTML}{2E86C1}
\definecolor{dark_blue}{HTML}{1B4F72}

\usepackage{hyperref}

\usepackage{orcidlink}

\begin{document}
\title{Short-distance thermal phase structure of charged black holes in 4D Einstein–Gauss–Bonnet gravity}

\author{Syed Masood\,\texorpdfstring{\orcidlink{0000-0003-3054-7317}}{}}
\email[Work: ]{masood@westlake.edu.cn}
\email[Personal: ]{quantummind137@gmail.com}
\thanks{Author full legal name (official records): Syed Masood Ahmad Shah Bukhari}
\affiliation{Zhejiang University/University of Illinois at Urbana-Champaign (ZJU-UIUC) Institute,
Zhejiang University, International Campus, Haining, Zhejiang 314400, China, \\}
\affiliation{Department of Physics, School of Science and Research Center for Industries of the Future, Westlake University, Hangzhou 310030, P. R. China,\\}
\affiliation{Institute of Natural Sciences, Westlake Institute for Advanced Study, Hangzhou 310024, P. R. China.\\}
\affiliation{Canadian Quantum Research Center, 204-3002 32 Ave, Vernon, BC V1T 2L7, Canada.\\}

\begin{abstract}
\textcolor{black}{Glavan and Lin’s proposal of an effective four-dimensional Einstein--Gauss--Bonnet (4D-EGB) gravity framework yields predictions that differ from general relativity in some regimes. A range of black hole studies have offered  insights  into the  dynamical and phenomenological aspects  of this effective theory of gravity.
In this work, the thermodynamics of a charged 4D-EGB black hole with  Gauss--Bonnet (GB) coupling $\alpha$, characterized by mass $M$ and charge $Q$ in the non-extremal regime $M>\sqrt{Q^2+\alpha}$ is investigated by combining a non-perturbative, quantum-gravity-inspired exponential correction to the entropy (quantified by $\eta$) with information-geometric diagnostics.
Within a canonical ensemble (fixed $Q$) paradigm, thermodynamic stability regions and phase-transition-like features are identified as the black hole size tends toward extremality due to Hawking evaporation.
The Ruppeiner metric is then constructed on the $(M,Q)$ state space and  the associated thermodynamic curvature is evaluated to characterize the effective interaction signatures and its relation to critical behavior.
In addition, an effective quantum-work quantity, defined from the free-energy landscape using Jarzynski equality, is evaluated as an additional probe of short-distance, near-extremal behavior.
The results indicate that departures from the general-relativistic behavior are negligible for large black holes but can become relevant at small horizon scales. Specifically, on short-distance scales, the combined influence of $\alpha$ and $\eta$ can modify stability  of the extremal black hole geometry and remnants within this thermodynamic model.}
\end{abstract}

\date{\today}
\maketitle

\section{Introduction}
\label{sec:intro}
\setcounter{page}{1}
\noindent \textcolor{black}{Black hole thermodynamics originates from the work of Hawking \cite{Hawking:1974rv, Hawking:1975vcx} and Bekenstein \cite{Bekenstein:1973ur} in semiclassical gravity, where entropy satisfies the area law,
\begin{equation}
S_{\rm BH}=\frac{A}{4\ell_{\rm p}^2},
\end{equation}
with $A$ the horizon area and $\ell_{\rm p}$ the Planck length \footnote{In this paper, the natural units $c=G=\hbar=1$ are used throughout.}
The area scaling is closely related to holography \cite{Susskind:1994vu, Bousso:2002hon}.
While this framework provides a consistent macroscopic description in terms of conserved charges, the microscopic origin of black hole entropy and the statistical interpretation of thermodynamic response functions remain open \cite{Davies:1978zz, Page:2004xp}}.

\textcolor{black}{Hawking evaporation \cite{Hawking:1974rv, Hawking:1975vcx} is derived in a semiclassical regime, and motivates corrections to the Bekenstein--Hawking description when the black hole approaches microscopic scales.
Many quantum-gravity approaches imply a minimum length and associated departures from classical geometry \cite{Hossenfelder:2012jw}, which in turn suggest modifications of the entropy \cite{Mann:1997hm, Upadhyay:2018vfu, Pourhassan:2019coq} and connections to microstate counting \cite{Strominger:1996sh}.
Entropy corrections predicted in various frameworks include perturbative (logarithmic or power-law) terms \cite{Rovelli:1996dv, PhysRevLett.80.904, PhysRevLett.84.5255, Dabholkar:2008zy, Mandal:2010cj} and non-perturbative contributions of exponential type \cite{Ashtekar:1991hf, Ghosh:2012jf, Dabholkar:2014ema, Chatterjee:2020iuf}.
Examples of the latter arise in AdS/CFT-based analyses \cite{Maldacena:1997re} and in treatments involving Kloosterman sums \cite{Murthy:2009dq, Dabholkar:2011ec, Dabholkar:2014ema}.
Such corrections are suppressed for macroscopic horizons but can dominate the thermodynamics near microscopic scales \cite{Dabholkar:2011ec, Dabholkar:2014ema, Chatterjee:2020iuf}; see Refs.~\cite{Hemming:2007yq, Gregory:2008br, Rocha:2008fe,Saraswat:2019npa,Mann:1997hm,Sen:2012cj,Govindarajan:2001ee,Pourhassan:2017qxi, Pourhassan:2017qhq, Pourhassan:2019coq, Pourhassan:2020bzu, Upadhyay:2019hyw, Ghaffarnejad:2022aqe, Biswas:2021gps,Pourhassan:2022sfk, Pourhassan:2022opb, Aounallah:2022rfo} for broader discussions.}

\textcolor{black}{A complete quantum-gravity treatment would modify both the geometry and thermodynamics of black holes.
Nevertheless, it is common in phenomenological studies to implement quantum effects through an effective entropy deformation while keeping the classical background fixed, thereby isolating their impact on response functions and stability criteria \cite{MasoodASBukhari:2023ljc, Pourhassan:2022sfk, Pourhassan:2022opb, Iso:2011gb}.
Since thermodynamic behavior depends on the underlying gravitational dynamics, it is also of interest to examine these effects in modified-gravity models.}

\textcolor{black}{General relativity (GR) is consistent with a wide range of observations, including gravitational waves \cite{LIGOScientific:2016aoc, LIGOScientific:2016lio} and black hole shadow measurements \cite{EventHorizonTelescope:2019dse}, but it also motivates extensions in regimes where curvature becomes large.
Higher-curvature theories provide a concrete class of such extensions \cite{Capozziello:2011et, Berti:2015itd, Shankaranarayanan:2022wbx}.
Gauss-Bonnet (GB) gravity, originating with Lanczos and Lovelock \cite{Lanczos:1938sf, Lovelock:1971yv, Lovelock:1972vz}, yields second-order field equations and avoids Ostrogradsky-type instabilities \cite{Ostrogradsky:1850fid}, and quadratic curvature terms appear naturally in string-inspired effective actions \cite{Zwiebach:1985uq,Gross:1986iv,Gross:1986mw}.
In four dimensions the GB invariant does not contribute to the classical field equations unless coupled to additional fields, such as scalars \cite{Blazquez-Salcedo:2016enn, Konoplya:2019fpy, Maselli:2014fca, Ayzenberg:2013wua}.
Glavan and Lin proposed a prescription leading to an effective four-dimensional Einstein-Gauss-Bonnet (4D-EGB) framework \cite{Glavan:2019inb}, which has been investigated in several directions including consistency analyses \cite{Lu:2020iav, Hennigar:2020lsl, Gurses:2020ofy}, quasinormal modes and shadows \cite{Konoplya:2020bxa, Kumar:2020owy}, and geodesic motion \cite{Guo:2020zmf}; see Ref.~\cite{Fernandes:2022zrq} for a review.} \\
\indent \textcolor{black}{Thermodynamic properties of 4D-EGB black holes  have been analyzed extensively, including extended thermodynamics and information-geometric diagnostics \cite{Wei:2020poh, HosseiniMansoori:2020yfj, EslamPanah:2020hoj,Hegde:2020yrd,Hegde:2020cdm,Hegde:2020cdm,Kumar:2020xvu,Kumar:2023ijg,Kumar:2024bls,Fernandes:2020rpa}.
Most existing studies of black hole thermodynamics adopt the conventional \textit{semiclassical} entropy  and focus on classical phase structure. For example, Ladghami \textit{et al.}~\cite{Ladghami:2023ccf}  considered charged black solutions in 4D-EGB gravity and provided a thermodynamic interpretation for GB coupling,  in addition to reporting several phase transitions contingent on the interplay between GB coupling and charge of the black holes. Moreover, the possibility of a logarithmic contribution (which is normally reported in quantum gravity paradigm \cite{Rovelli:1996dv, PhysRevLett.80.904, PhysRevLett.84.5255, Dabholkar:2008zy, Mandal:2010cj}) beyond the leading Bekenstein-Hawking term to the  black hole entropy has been reported by Fernandes \cite{Fernandes:2020rpa}. Wei and Liu \cite{Wei:2020poh} applied Ruppeiner information geometry to charged 4D‑EGB black holes, identifying van‑der‑Waals–type small/large black hole phase behavior, and using the thermodynamic curvature to diagnose regimes suggestive of attractive/repulsive effective interactions. A similar investigation by Mansoori \cite{HosseiniMansoori:2020yfj} utilizes a novel information geometric probe \cite{HosseiniMansoori:2019jcs} to  quantify the role of GB coupling on black hole microstructures within 4D-EGB gravity. Several other studies that cover various aspects of charged black holes in 4D-EGB gravity can be found in refs.\cite{Yang:2020czk,Yang:2025xck,Guo:2022kio,Kumar:2020sag}. Whether the aforementioned thermal behavior in 4D‑EGB gravity changes qualitatively once quantum-scale (short-distance) effects become relevant remains an open  question. This  motivates us to examine how non-perturbative, short-distance quantum corrections to the entropy reshape stability criteria and information-geometric signatures within the 4D‑EGB landscape.}\\
\indent \textcolor{black}{In this work, we consider a charged 4D-EGB black hole in the GR branch and restrict to the non-extremal regime $M>\sqrt{Q^2+\alpha}$ in a fixed-$Q$ (canonical) setup.
Quantum effects are incorporated through a non-perturbative exponential deformation of the entropy \cite{Chatterjee:2020iuf}, while the geometry and the Hawking temperature defined by surface gravity are kept classical, so the results should be interpreted as effective thermodynamic diagnostics \cite{MasoodASBukhari:2023ljc}.
Our main deliverables are: (i) heat capacity and Helmholtz free-energy analyses of stability and phase-transition-like behavior; (ii) an effective quantum-work characterization based on Jarzynski equality; and (iii) the Ruppeiner thermodynamic geometry \cite{Ruppeiner:2013yca} on the $(M,Q)$ state space, where curvature signatures are correlated with heat-capacity structure. The investigations find relevance to extremality and the associated black hole remnants in quantum regime. To our knowledge, this is the first 4D‑EGB information‑geometric study that combines exponential (non‑perturbative) entropy  and Jarzynski-based work diagnostics in the near‑extremal regime with relevance to quantum domain.}\\
\indent \textcolor{black}{This article is organized as follows.
Section~\ref{sec:concepts} reviews the 4D-EGB black hole solution and the entropy deformation.
Section~\ref{sec:heatcapacity} analyzes the heat capacity and stability.
Section~\ref{sec:QWD} discusses the quantum-work diagnostic.
Section~\ref{sec:Ruppgeometry} presents the thermodynamic geometry analysis.
Section~\ref{sec:summary} summarizes our work.}

\section{\label{sec:concepts}
Conceptual aspects}
     \subsection{\textcolor{black}{4D-EGB gravity and dimensional regularization}}
  \noindent   In standard GR, the 4D Einstein-Hilbert action  is written as 
   \begin{eqnarray}
       S_{\rm EH}=\int \mathrm{d}^4x\sqrt{-g}R,
   \end{eqnarray}
   where $R$ is the Ricci curvature scalar, and $g$ the determinant of metric tensor $g_{\mu\nu}$. Lovelock theorem \cite{Lovelock:1971yv, Lovelock:1972vz, Lanczos:1938sf} asserts that GR is the unique theory of gravity in four dimensions, provided certain conditions are met. These conditions include diffeomorphism invariance, metricity, and second-order equations of motion. In higher-dimensional spacetime, the action that satisfies these three conditions is the GB action, given by
   \begin{eqnarray}\label{actionGB}
       S_{\rm GB}=\int \mathrm{d}^{D}x\sqrt{-g}\left(R+\alpha \mathcal{G}\right)
   \end{eqnarray}
   where $ \mathcal{G}=R^2-4R_{\mu\nu}R^{\mu\nu}+R_{\mu\nu\rho\sigma}R^{\mu\nu\rho\sigma}$ is the GB invariant. Here, $R^{\mu\nu}$ and $R^{\mu\nu\rho\sigma}$ are the Ricci and Riemann tensors, respectively. By varying the action in Eq. (\ref{actionGB}) with respect to metric tensor $g_{\mu\nu}$, we get the following field equations of gravity:
   \begin{eqnarray}
      R_{\mu\nu}-\frac{1}{2}Rg_{\mu\nu}+\alpha H_{\mu\nu}=T_{\mu\nu},
   \end{eqnarray}
   where \begin{align}\nonumber
H_{\mu\nu}=&2RR^{\mu\nu}-4R_{\mu\sigma}R^{\sigma}_{\nu}-4R_{\mu\sigma\nu\rho}R^{\sigma\rho}\\ 
       &-2R_{\mu\sigma\nu\delta}R^{\sigma\rho\delta}_{\nu}-\frac{1}{2}\mathcal{G}g_{\mu\nu},
  \end{align}
   and $T_{\mu\nu}$ is the stress-energy tensor. Until recently, it was well known that in ordinary 4D spacetime, the fact that $\mathcal{G}$ is a total derivative implies it has no contribution to the dynamics. However, the idea proposed in Ref. \cite{Glavan:2019inb} suggested a way to circumvent this conclusion through a novel rescaling of the coupling constant by
   \begin{eqnarray}\label{GLP}
       \alpha\rightarrow \frac{\alpha}{D-4},
   \end{eqnarray}
which, after taking the limit $D\rightarrow 4$, produces nontrivial contributions of the GB term to the dynamics. \\
\indent  \textcolor{black}{
At the level of the equations of motion, the Glavan--Lin rescaling prescription (\ref{GLP}) has been the subject of substantial discussion.
In particular, it has been emphasized that taking $D\!\to\!4$ is not a standard continuous limit for tensorial expressions and can lead to ambiguities beyond highly symmetric ans\"atze \cite{Ai:2020peo}, and that the naive construction does not generically define consistent four-dimensional field equations without additional prescriptions \cite{Gurses:2020rxb}.
Related analyses have further argued that, even when linearized fluctuations around maximally symmetric backgrounds are well behaved, the perturbative expansion becomes ill-defined at higher order, indicating that additional structure is required to obtain a well-defined four-dimensional theory \cite{Arrechea:2020evj}.
Motivated by these issues, several works have proposed explicit $D\!\to\!4$ completions/regularizations---typically involving an additional scalar-type degree of freedom and/or a controlled breaking of diffeomorphism invariance---which aim to provide well-defined four-dimensional field equations   
\cite{Aoki:2020lig,Lu:2020iav,Hennigar:2020lsl,Gurses:2020ofy,Charmousis:2025jpx}, and phenomenological/observational implications  \cite{Charmousis:2021npl,Garcia-Aspeitia:2020uwq,Wang:2021kuw,Khodabakhshi:2024gpo}. Related discussions and other broader perspectives of the theory can be found in the review work \cite{Fernandes:2022zrq}.
Since there is no unique universally adopted resolution of the dimensional regularization problem, in this work, we employ the spherically symmetric black hole solution arising from the Glavan--Lin approach as an effective phenomenological background and interpret the thermodynamic results within this specific 4D-EGB setting.
We adopt this prescription here because it provides a widely used, closed-form static black hole background that enables direct comparison with earlier thermodynamic studies in the 4D‑EGB literature. In the present work, the rescaling affects our results only through the chosen metric function (and hence the horizon structure/extremality bound), on top of which we implement an exponential entropy deformation while keeping the geometry classical. The conclusions should therefore be interpreted as model-dependent thermodynamic diagnostics within this specific effective 4D‑EGB construction. We thus expect that alternative $D\rightarrow 4$ completions could lead to different quantitative behavior.}

\subsection{Black holes in 4D-GB theory}
The novel theory of 4D-GB gravity thus predicts a line element of the form \cite{Glavan:2019inb,Fernandes:2022zrq}
\begin{eqnarray}
    \mathrm{d}s^2=-f(r)\mathrm{d}t^2+\frac{1}{f(r)}\mathrm{d}r^2+r^2\left(\mathrm{d}\theta^2+\sin^2{\theta}\mathrm{d}\phi^2\right)
\end{eqnarray}
with
\begin{equation}\label{fr}
f(r)=1+\frac{r^2}{2\alpha}\left[1\pm\sqrt{1+\left(\frac{8\alpha M}{r^3}\right)}\right].
\end{equation} 
There are two branches of solutions for the above metric: the plus sign indicates the GB branch, while the minus sign represents the GR branch. The physical viability of these branches can be established by examining the limits of $\alpha$ and $r$ in the solutions.  \textcolor{black}{Let's first consider the far-field limit $r \rightarrow \infty$, for which GR branch from Eq. (\ref{fr}) yields the following:}
\begin{eqnarray}
    f(r)=1-\frac{2M}{r}+\mathcal{O}\left(\frac{1}{r^2}\right).
\end{eqnarray}
\textcolor{black}{which is the usual Schwarzschild solution.} Now for the GB branch, we have 
\begin{eqnarray}
    f(r)=1+\frac{2M}{r}+\frac{r^2}{\alpha}+\mathcal{O}\left(\frac{1}{r^2}\right),
\end{eqnarray}
\textcolor{black}{which is not asymptotically flat, i.e., does not recover the expected GR limit, and  is therefore excluded from the analysis}. Furthermore, the condition $\alpha \rightarrow 0$ for the GR branch gives a well-defined limit
\begin{eqnarray}
    f(r)=1-\frac{2M}{r}+\mathcal{O}\left(\alpha\right),
\end{eqnarray}
while for GB branch, one obtains 
\begin{eqnarray}
    f(r)=1+\frac{2M}{r}+\frac{r^2}{\alpha}+\mathcal{O}\left(\alpha\right),
\end{eqnarray}
which is not a well-defined limit. Note also that the mass term for the GB branch has the wrong sign \cite{Glavan:2019inb,Fernandes:2022zrq}. Hence, this branch is discarded as a viable physical solution. On these grounds, we only consider the GR branch in our analysis. 
\begin{figure*}[t]
\centering
\includegraphics[width=0.8\textwidth, height=6cm]{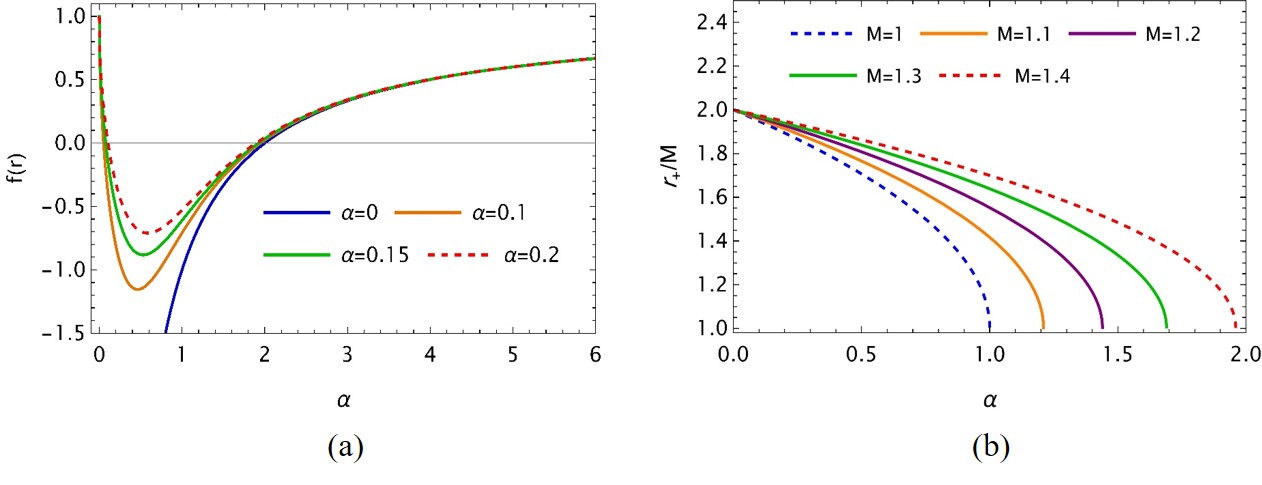}
\caption{Impact of $\alpha$ on (a) the metric function  $f(r)$, where it indicates finiteness of the metric coefficient at $r\rightarrow 0$ compared to Einstein gravity ($\alpha=0$) , and (b) horizon radius $r_{+}$ depicting how $\alpha$ shrinks the black hole.}%
\label{frrgplot}%
\end{figure*}
Next, we investigate the impact of $\alpha$ on the metric function and horizon radius. Solving Eq. (\ref{fr}) for $r$ yields
\begin{eqnarray}\label{rpm1}
    r_{\pm}=M\pm\sqrt{M^2-\alpha},
\end{eqnarray}
where $r_{+}$ represents the event horizon radius of the black hole, while $r_{-}$ is the Cauchy horizon. For $\alpha\rightarrow 0$, this yields the usual Schwarzschild solution. 

We graphically illustrate the metric function $f(r)$ and the horizon radius $r_{+}$ against GB parameter 
$\alpha$ in Fig.\ref{frrgplot}. It can be readily observed from Fig. \ref{frrgplot}(a) that $f(r)$ for the 4D-GB black hole is finite near the origin $r \rightarrow 0$. However, the metric coefficients being finite at the origin $r \rightarrow 0$ do not eliminate the central singularity, as the Kretschmann curvature scales as $R_{\mu\nu\rho\sigma}R^{\mu\nu\rho\sigma} \propto r^{-3}$ \cite{Fernandes:2022zrq}.
Interestingly, in GR for a Schwarzschild black hole, the Kretschmann scalar scales as $r^{-6}$ at any radius $r$. This implies that the introduction of the GB coupling parameter $\alpha$ weakens the black hole singularity. One can also see from Fig. \ref{frrgplot}(b) that for a given black hole mass $M$, the introduction of $\alpha$ reduces the black hole size compared to the GR case. This subtle effect arises purely from higher curvature corrections to Einstein gravity, providing a physical picture of the nontrivial contributions of $\alpha$ to gravitational dynamics. Additionally, we note the presence of another singularity at a radius where $ r^3 = -8\alpha M $, rendering the expression under the square root in Eq. (\ref{fr}) zero. While the parameter $\alpha$ is permitted to take both positive and negative values \cite{Guo:2020zmf,Konoplya:2020bxa}, its negative values are subject to stringent constraints \cite{Charmousis:2021npl}. In this work, we focus exclusively on positive values of $\alpha$, following the approach in the original study \cite{Glavan:2019inb}.

\begin{figure}
\centering
{\includegraphics[width=7.5cm, height=5.5cm]{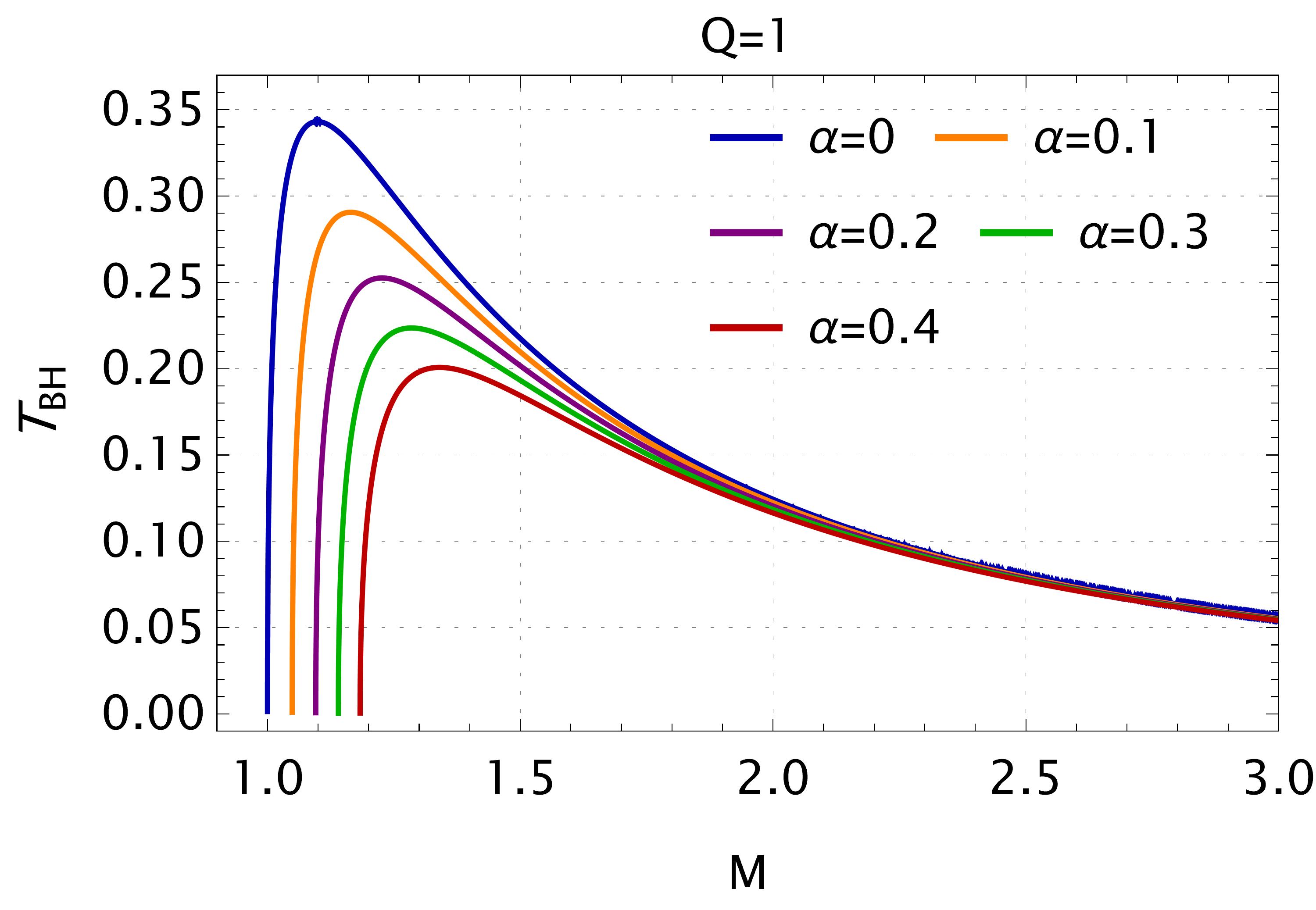}}
\caption{Black hole temperature $T_{\rm BH}$ versus the mass $M$. The GB coupling $\alpha$ parameter tends to make the black hole colder on smaller scales, mimicking the role of charge in \textcolor{black}{RN geometry}\label{Tplot}.}
\end{figure}

\indent  Charged black hole solutions in 4D-GB gravity have also been found, for which the following relation holds \cite{Fernandes:2020rpa}
\begin{equation}\label{frQ}
f(r)=1+\frac{r^2}{2\alpha}\left[1-\sqrt{1+4\alpha\left(\frac{2M}{r^3}-\frac{Q^2}{r^4}\right)}\right].
\end{equation}  
For the non-extremal geometry scenario, under which $M>\sqrt{Q^2+\alpha}$, the zeros of $f(r)$ give rise to two horizons located at    
\begin{equation}\label{root}
r_{\pm}=M\pm\sqrt{M^{2}-Q^{2}-\alpha},
\end{equation}
where $r_{+}$ is the black hole horizon, and $r_{-}$ the inner Cauchy horizon. One can work with a timelike Killing vector $\xi^{\mu}$, which allows us to define the surface gravity $\kappa$ of the black hole as $\kappa = \nabla_{\mu} \xi^{\mu} \nabla_{\nu} \xi^{\nu} = 1/2 \left[ \mathrm{d}f(r)/\mathrm{d}r \right]_{r = r_{+}}$. This surface gravity corresponds to a black hole temperature of \textcolor{black}{$\kappa/2\pi$}. Utilizing $r_{+}$ from Eq. (\ref{root}), this yields:
\begin{widetext}
\begin{eqnarray}\label{temp1}
T_{\rm BH}= \frac{\left(M+\sqrt{M^2-Q^2-\alpha}\right)^3 \left[\sqrt{\frac{\left(M+\sqrt{M^2-Q^2-\alpha}\right)^4+8 \alpha M \left(M+\sqrt{M^2-Q^2-\alpha}\right)-4\alpha Q^2}{\left(M+\sqrt{M^2-Q^2-\alpha}\right)^4}}-1\right]-2 \alpha M}{\alpha \left(M+\sqrt{M^2-Q^2-\alpha}\right)^2
   \sqrt{\frac{\left(M+\sqrt{M^2-Q^2-\alpha}\right)^4+8 \alpha M \left(M+\sqrt{M^2-Q^2-\alpha}\right)-4 \alpha Q^2}{\left(M+\sqrt{M^2-Q^2-\alpha}\right)^4}}}.
\end{eqnarray}
\end{widetext}
The quantity $T_{\rm BH}$ is plotted in Fig. \ref{Tplot}, illustrating that the GB coupling parameter $\alpha$ reduces the black hole temperature at smaller scales, exhibiting a behavior similar to that of the charge $Q$ in \textcolor{black}{Reissner-Nordstr\"om (RN) geometry} . Notably, we did not introduce any corrections to $T_{\rm BH}$ due to $\eta$ (see Eq. \eqref{S_np}). The reason for this is that we use a modified entropy formula based on quantum gravity principles \cite{Chatterjee:2020iuf} without invoking any quantum geometry. Consequently, the metric function $f(r)$ in Eq. (\ref{fr}) remains unchanged, and the definition of temperature $(T_{\rm BH} \propto \mathrm{d}f(r)/\mathrm{d}r)$ follows from this function. This approach has been discussed in some earlier works \cite{Pourhassan:2022sfk,Pourhassan:2022opb,MasoodASBukhari:2023ljc}.
\subsection{Entropy in the quantum regime}
One of the earliest attempts to understand the microscopic degrees of freedom for black hole entropy originates from the string theory approach \cite{Strominger:1996sh}. However, as previously mentioned, approaches to quantum gravity and string theory through microstate counting predict additional subleading perturbative or non-perturbative contributions to the original Bekenstein-Hawking term. Nevertheless, all these approaches yield only the Bekenstein-Hawking contribution for larger geometries, with the extra terms becoming significant only in the quantum regime. A more robust and fundamental approach would be to quantize the gravitational action and deduce the resulting thermodynamic behavior. However, this task is exceedingly difficult due to the mathematical complexity involved and the uncertainty about the ultimate physical assumptions underlying quantum gravity. Yet one can adopt a more pragmatic approach by considering only the quantum corrections to entropy, thereby exploring black hole thermodynamics in the quantum regime. In this context, the Jacobson framework \cite{PhysRevLett.75.1260} and its connections to thermal fluctuations \cite{Faizal:2017drd} may serve as a motivational basis. 

\indent The perturbative contributions to the quantum corrections of black hole entropy have the following general form \cite{PhysRevLett.80.904, PhysRevLett.84.5255,Pourhassan:2017qhq,Pourhassan:2019coq}:
\begin{eqnarray}
 S_{\rm p}=\mathcal{A}\ln{\left(\frac{A}{4\ell_{\rm p}^2}\right)}+\frac{ 4\mathcal{B} \ell_{\rm p}^2}{A}+\cdots,
\end{eqnarray}
where as before $A$ denotes horizon area of the black hole while $\mathcal{A}$ and $\mathcal{B}$ are some constants related to the quantum gravity scale. Now the non-perturbative corrections read as \cite{Dabholkar:2011ec, Dabholkar:2014ema, Chatterjee:2020iuf}:
\begin{eqnarray}\label{S_np}
 S_{\rm np}=\eta e^{-A/4\ell_{\rm p}^2}.
\end{eqnarray}
Here, $\eta$ is a positive parameter measuring the scale of the non-perturbative contribution to the black hole's entropy.
With that, the total BH entropy of the black hole would be $S_{\rm BH}=S_{0}+S_{\rm p}+S_{\rm np}$. There is an intriguing aspect to black holes in 4D-GB theory: the entropy already includes logarithmic contributions (albeit perturbative) from classical geometry considerations, as outlined in Ref. \cite{Fernandes:2020rpa}. Taking note of this, after incorporating exponential corrections \cite{Chatterjee:2020iuf}, we express the total entropy of the black hole as
\begin{eqnarray}\label{S1}
    S_{\rm exp}=S_{0}+\eta e^{-S_{0}},
\end{eqnarray}
where $S_{0}=A+\alpha \log{A}$ is the original entropy reported in Ref. \cite{Fernandes:2020rpa}. 
\begin{widetext}
Given that $A=r_{+}^2$ (omitting proportionality constants), we expand Eq. (\ref{S1}) as follows:  
\begin{align}\nonumber
 S_{\rm exp}&=S_{0}+\eta \exp{\left(-S_{0}\right)}\\[4pt]
 \nonumber
 &=\left(M+ \sqrt{M^2-Q^2-\alpha}\right)^2+2\alpha \log{\left(M+ \sqrt{M^2-Q^2-\alpha}\right)}\\[4pt]
& +\eta \exp{\left[-\left(M+ \sqrt{M^2-Q^2-\alpha}\right)^2-2\alpha \log{\left(M+ \sqrt{M^2-Q^2-\alpha}\right)}\right]}.
 \end{align}
\end{widetext}

\begin{figure*}[t]
\centering
\includegraphics[width=0.8\textwidth, height=6cm]{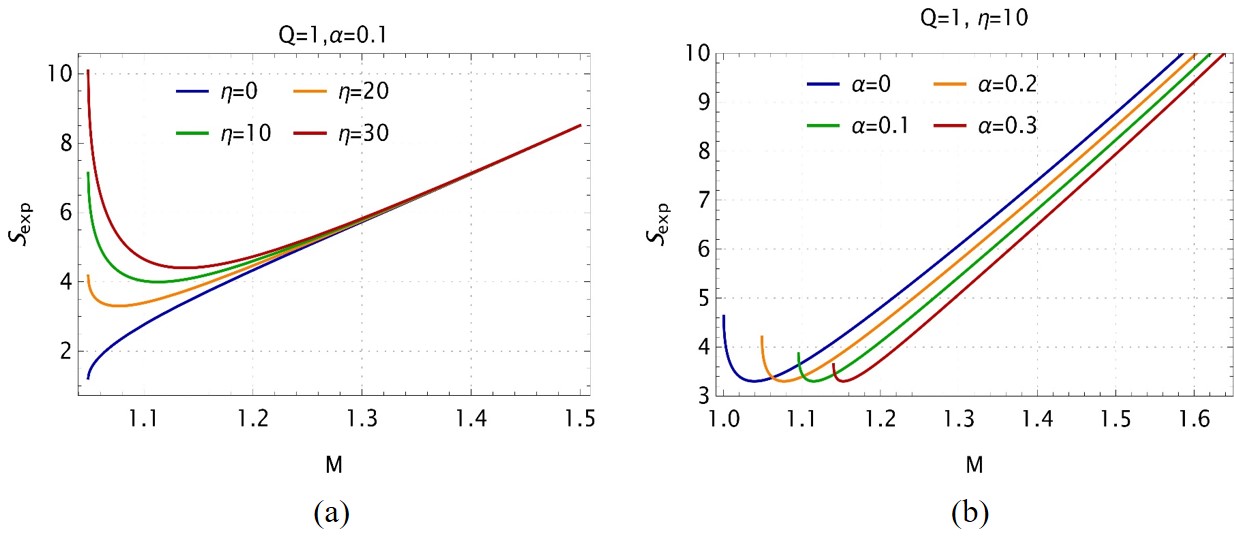}
\caption{Variation of black hole entropy with respect to (a) exponential parameter $\eta$, and (b) GB coupling $\alpha$. Exponential contributions become dominant on quantum scales compared to Bekenstein-Hawking term.}%
\label{entropy}%
\end{figure*}

\indent Fig. \ref{entropy} illustrates the distinctive nature of the entropy curves corresponding to different values of $\alpha$ (GB parameter) and $\eta$ (the quantum correction scale). For larger sizes, all curves tend to coincide, reflecting the dominance of the Bekenstein-Hawking term. It's important to note that our black hole system features bifurcate horizons and exhibits distinct non-extremal and extremal geometric descriptions corresponding to the cases $M > \sqrt{Q^2 + \alpha}$ and $M = \sqrt{Q^2 + \alpha}$, respectively. Beyond the extremal limit, the black hole singularity becomes naked, a scenario typically forbidden by the cosmic censorship conjecture \cite{Penrose:1969pc}. Due to the non-perturbative nature of exponential corrections in the quantum regime, the plots exhibit a sudden jump near the extremal limit whenever $M = \sqrt{Q^2 + \alpha}$. It's noteworthy that the entropy reaches a large value at this point but does not diverge.   
Upon further examination of the plots, we observe that $\eta$ contributes to a larger entropy [Fig. \ref{entropy}(a)], whereas $\alpha$ tends to hasten the end of evaporation (for a fixed $Q$) by shifting the extremal limit $M = \sqrt{Q^2 + \alpha}$ each time $\alpha$ is changed [Fig. \ref{entropy}(b)]. Additionally, from Fig. \ref{entropy}(b), we notice that for the $\alpha=0$ case, the jump in entropy occurs precisely at $M=Q$, a characteristic of the \textcolor{black}{RN black holes} in GR.\\
\indent As quantum corrections to the entropy become relevant near the extremal limit $M=\sqrt{Q^2+\alpha}$, it is suggestive of the black hole geometric scales and the applicability of quantum corrections to the entropy. The conventional perspective of Hawking evaporation assumes that the diminishing black hole size involves the entirety of its horizon radius $r_{+}$, which naturally encompasses a combination of all three parameters in our case: $M$, $Q$, and $\alpha$.
This fact is well-established concerning charged black holes \cite{Hiscock:1990ex}, where the charge-to-mass ratio evolves as the black hole continues its evaporation towards the extremal limit. However, our definition assumes a canonical ensemble framework where $M$ fluctuates while $Q$ and $\alpha$ remain constant. This implies that the entire evaporation process occurs via $M$, which governs the black hole's geometric scales. However, it's important to exercise caution regarding the magnitudes of all three parameters $M$, $Q$, and $\alpha$, as their values must satisfy the quantum gravity scale in relation to entropy corrections. Here, we enforce the condition that $Q$ and $\alpha$ are extremely small, providing nearly negligible contributions. Consequently, whenever $M \rightarrow \sqrt{Q^2 + \alpha}$, i.e., the extremal limit, our black hole tends to possess a quantum description; otherwise, it behaves as a classical system for all $M > \sqrt{Q^2 + \alpha}$.
This description would place the black hole in a coexisting phase of classical and quantum descriptions corresponding to a characteristic value of $M$. As we will later observe, this particular value of $M$ corresponds to the first root of the heat capacity $C_{Q}$ (Fig. \ref{CQ}). With that said, the quantum corrections to the entropy via Eq. (\ref{S1}) naturally follow.

\section{\label{sec:heatcapacity} Black hole stability via heat capacity}
The thermodynamic stability of black holes can be explored through the examination of various thermodynamic potentials, depending on the chosen ensemble approach. \textcolor{black}{For instance, in the canonical ensemble (fixed $Q$) approach, one can define the heat capacity $C_{Q}$ of the system by the following expression:}
\begin{eqnarray}\label{C_Q}
    C_{Q}=T\left(\frac{\partial S_{\rm exp}}{\partial T}\right)_{Q}=T\left(\frac{\partial S_{\rm exp}/\partial M}{\partial T/\partial M}\right)_{Q}.
\end{eqnarray}

\begin{figure*}[t]
\centering
\includegraphics[width=\textwidth, height=5.4cm]{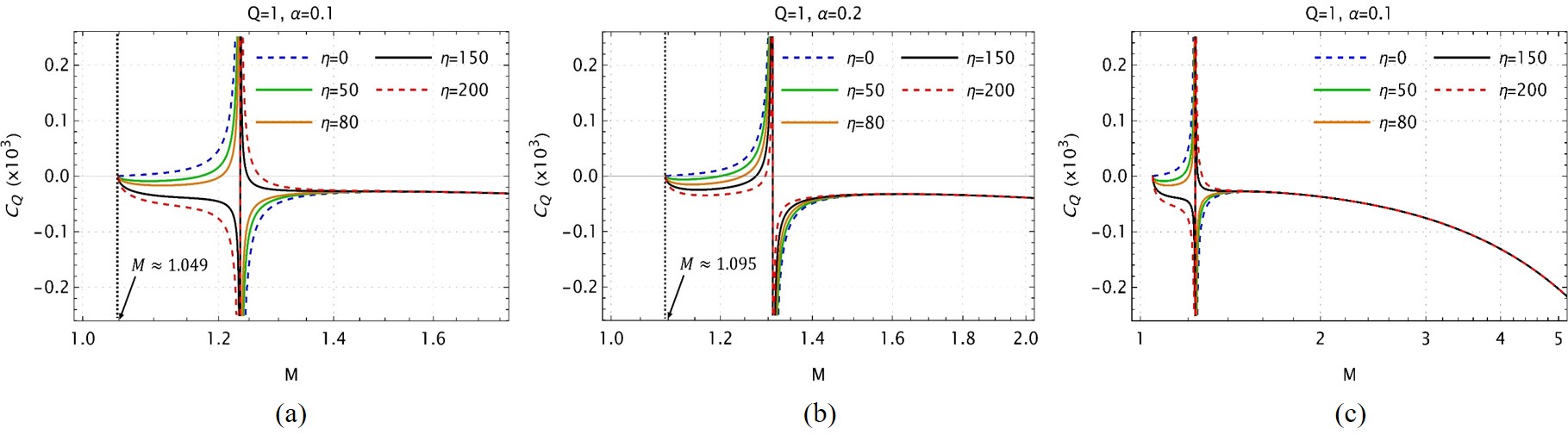}
\caption{Heat capacity $C_{Q}$ as a function of black hole mass $M$ for $Q=1$ and GB coupling (a) $\alpha=0.1$ and (b) $\alpha=0.2$. The second zero of $C_{Q}$ occurs whenever $M=\sqrt{Q^2+\alpha}$.}%
\label{CQ}%
\end{figure*}

Substituting Eq. (\ref{S1}) into Eq. \eqref{C_Q}, the following formula is obtained:
\begin{widetext}
\begin{align}\nonumber
C_{Q}&= \frac{2 e^{-r_{+}^2} \left(Q^2+\alpha-2 M r_{+}-1\right)
   \left(-r_{+}^4-8 \alpha M r_{+}+4 \alpha Q\right) \left(\mathcal{Y}-2 \alpha M\right)
   \left[e^{r_{+}^2} \left(Q^2+\alpha-2 M r_{+}\right)+\eta
   \right]}
      {r_{+}^2 \left[\mathcal{A+B +C  +D}+2 \left(-r_{+}^4-8
   \alpha M r_{+}+4 \alpha Q\right) \left(2 \alpha M-Y\right)\right]},
\end{align}
where, for brevity, we introduce the quantities
\begin{align*}
\mathcal{Y}=r_{+}^3\left(\sqrt{\frac{8 \alpha M r_{+}-4 \alpha Q}{r_{+}^4}+1}-1\right);\qquad
\mathcal{A}=3 \left(r_{+}^4+8 \alpha M r_{+}-4
   \alpha Q\right) \left(1-\sqrt{\frac{8 \alpha M r_{+}-4 \alpha Q}{r_{+}^4}+1}\right)
   r_{+}^3,\\
   \mathcal{B}=4 \alpha r_{+}^3\left[\alpha+(Q-2) Q+2 M r_{+}\right] \sqrt{\frac{8 \alpha M
   r_{+}-4 \alpha Q}{r_{+}^4}+1},\qquad
\mathcal{C}=2 \alpha\sqrt{M^2-Q^2-\alpha} \left(r_{+}^4+8 \alpha M r_{+}-4 \alpha Q\right), \\
\mathcal{D}= 4 \alpha
\left[\alpha+(Q-2) Q+2 M r_{+}\right] \left[2 \alpha M-r_{+}^3 \left(\sqrt{\frac{8 \alpha M
   r_{+}-4 \alpha Q}{r_{+}^4}+1}-1\right)\right].
\end{align*}
   \end{widetext}

\noindent The  heat capacity is  displayed in Fig. \ref{CQ}. It's important to note that a negative heat capacity indicates an unstable thermodynamic phase, and vice versa \cite{Davies:1978zz}. \textcolor{black}{We observe from the plots that regardless of  $\alpha$ and  $\eta$,  $C_{Q}$ is negative for this charged black hole on larger scales [Fig. \ref{CQ}(c)].}
The scenario unfolds differently as the black hole geometry shrinks due to Hawking evaporation. $C_{Q}$ either tends to become more negative [Fig. \ref{CQ}(a), $\eta=0,50$, and $80$], or transitions to positive values via $C_{Q}=0$ [Fig. \ref{CQ}(a), $\eta=150$, and $200$] before encountering an infinite discontinuity at a characteristic mass value $M$. This distinct behavior of $C_{Q}$ arises for specific choices of $\eta$ and $\alpha$, and may be absent in other cases [see Fig. \ref{CQ}(b)]. We also observe that compared to the original uncorrected case ($\eta=0$), which indicates a single root for $C_{Q}$, all curves with $\eta \neq 0$ possess two zeros of $C_{Q}$. The transition of $C_{Q}$ from negative to positive values indicates that the black hole phase changes from being unstable to a stable one, reflecting a second-order phase transition in charged black holes \cite{Davies:1978zz}. This transition closely resembles familiar thermodynamic phase transitions in non-gravitational systems, such as ferromagnetic to paramagnetic, conductor to superconductor, liquid-crystal phase transitions, etc.\\
\indent  The zeros of $C_{Q}$ are typically interpreted as critical points that help distinguish between positive and negative  temperature solutions \cite{Hendi:2018sbe}. In this case, the second zero of $C_{Q}$ occurs at the extremal limit where black hole evaporation ceases, corresponding to $M = \sqrt{Q^2 + \alpha}$. The first zero of $C_{Q}$ marks the onset of the phase transitions and represents the point where our earlier  discussion regarding the definition of classical and quantum geometry for the black hole, in relation to entropy corrections, becomes relevant.
This endpoint related to $M = \sqrt{Q^2 + \alpha}$ may signify the formation of a black hole remnant, and is independent of $\eta$ in our case \textcolor{black}{(our horizon geometry is $\eta$-independent)}. However, it manifests explicit dependence on $\alpha$, which we have numerically computed and indicated in terms of certain characteristic values of $M$ for $\alpha=0.1$ and $0.2$ in Figs. \ref{CQ}(a) and (b), respectively. Each time $\alpha$ takes new values, the extremal limit and  corresponding shift in the last root of $C_{Q}$ is observed. \\
\indent \textcolor{black}{Compared to the general-relativistic RN baseline $(\alpha=0)$, where the extremal (zero-temperature) endpoint occurs at $M=|Q|$, the effective 4D-EGB background employed here simply shifts the extremality bound to
$M=\sqrt{Q^2+\alpha}$.
For fixed $Q$ and $\alpha>0$, this implies a larger terminal mass scale; equivalently, in a fixed-$Q$ evaporation picture, the evolution reaches $T_{\rm H}\to 0$ limit after a smaller decrease in $M$ than in GR, yielding a heavier remnant within this effective thermodynamic model.
This GR comparison should be interpreted as a consequence of the modified horizon/extremality condition in the chosen background, while quantum effects are incorporated here only through the entropy deformation and without modeling charge loss or backreaction explicitly. Meanwhile, determining the (in)stability of this remnant from $C_{Q}$ alone is a bit challenging , as we only explore up to the extremal limit of the black hole geometry. However, we will demonstrate in Section \ref{sec:Ruppgeometry} that thermodynamic geometry offers a slightly better way to address this issue.} \\

\section{\label{sec:QWD} Quantum work distribution}
\begin{figure*}[t]
\centering
\includegraphics[width=0.8\textwidth, height=6cm]{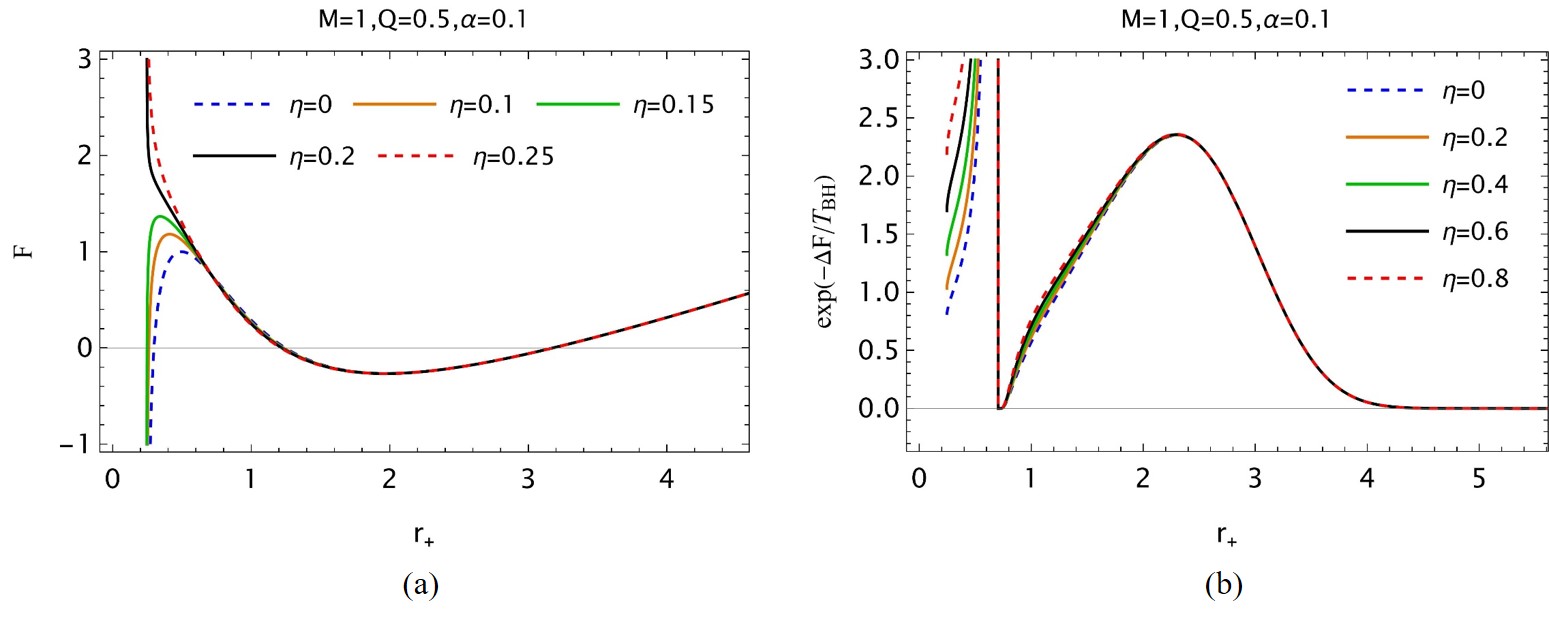}
\caption{\textcolor{black}{(a) Impact of quantum corrections on the Helmholtz free energy of the black hole, and (b) Quantum work as a function of the black hole size, where it is significant for microscopic geometries and negligible for larger scales.}}%
\label{QW}%
\end{figure*}
The thermodynamic behavior of black holes suggests the existence of a possible microstate structure. Each time a black hole emits radiation quanta due to Hawking evaporation, it alters the count of its microstates. As the black hole approaches microscopic scales, possibly near the Planck scale, quantum corrections and spacetime fluctuations become dominant. These effects manifest in the geometry of the black hole and in the associated Bekenstein-Hawking thermodynamics, as demonstrated in the preceding discussion.
In this context, classical thermodynamic potentials, such as entropy and free energy, are modified due to quantum corrections. These modifications have significant implications for the partition functions and the fluctuating microstates that model black hole evaporation. In this framework, quantum work serves as a valuable quantity, quantifying the relative probabilities of transitions in a quantum system in terms of its free energy.\footnote{Thus, the concept of quantum work has become a central tool in understanding quantum thermodynamics, a field grounded in non-equilibrium thermodynamic principles \cite{Fei:2020bjh,Wei:2017jvt,Salmilehto:2014xxa}.} 

Quantum work has been applied to study fluctuating black hole states using a stochastic approach based on Langevin and Fokker-Planck equations, as well as Jarzynski equality \cite{Iso:2010tz,Iso:2011gb,Pourhassan:2021mhb,Soroushfar:2023acx,Pourhassan:2022sfk}. 
For macroscopic black holes, the fluctuations are negligible as the geometry is effectively classical. However, on microscopic scales, quantum corrections to entropy become significant, making quantum work a crucial factor in the thermodynamic description of the black hole. \textcolor{black}{ More important of these quantum corrections are the ones that manifest as exponential (non-perturbative) contributions~\cite{Chatterjee:2020iuf}.} Consequently, our goal here is to investigate the effect of non-perturbative  corrections on the quantum work of the black hole. To do this, we first compute the Helmholtz free energy of the black hole. From Eq. \ref{frQ}, we can solve for $ M $ in terms of $ r_{+} $ to obtain
\begin{eqnarray}
    M=\frac{r_{+}^2+Q^2+\alpha}{2 r_{+}}.
\end{eqnarray}
This expression represents the internal energy of the black hole. Furthermore, we can rewrite the expression for temperature (Eq. \ref{temp1}), leaving out constant factors, as follows:
\begin{eqnarray}
    T_{\rm BH}=\frac{r_{+}^3 \left(\mathcal{K}-1\right)-2 \alpha  M}{\alpha  r_{+}^2 \mathcal{K}},
\end{eqnarray}
where $\mathcal{K}:=\sqrt{1-\frac{4 \alpha  (Q^2-2 M r_{+})}{r_{+}^4}}$.
 Helmholtz free energy of the black hole is given by
\begin{widetext}
\begin{align}\nonumber
    F=M-T_{\rm BH}S=\frac{\alpha +Q^2+r_{+}^2}{2 r_{+}}
    -\frac{\left(r_{+}^3 \left[\mathcal{K}-1\right]-2 \alpha  M\right) \left(r_{+}^2+\eta  e^{-r_{+}^2} r_{+}^{-2 \alpha }+2 \alpha  \log r_{+}\right)}{\alpha  r_{+}^2 \mathcal{K}},
\end{align}
\end{widetext}
which has been  plotted in Fig.\ref{QW}(a) for various values of $\eta$. The plots for $F$ reveal distinct features across various size ranges. High values of $F$ for larger black hole geometric scales indicate thermodynamic instability, which is not surprising given the overall character of the black hole, particularly its larger size, which resembles the \textcolor{black}{RN-like behavior noted earlier via $C_{Q}$ analysis (see Fig.\ref{CQ}).} The divergence of $F$ as $r_{+} \to 0$ reflects the true singular nature of the black hole at this limit. However, one would not expect non-zero values of $F$ at this point on physical grounds, as Hawking evaporation has removed all matter from the black hole. Thus, it is essential to interpret this as a mathematical artifact arising from the geometric properties of the black hole from which thermodynamics is defined. Clearly, the singularity near $r_{+} = 0$ is a critical feature, where the geometry becomes singular. Additionally, the black hole undergoes a transition through a region of thermodynamic stability, marked by negative free energy, as it evaporates from larger unstable configurations toward the divergent $F$ at $r_{+} = 0$. These stable/unstable phase transitions are a common feature in charged black hole thermodynamics, resulting from the competition between charge and mass \cite{Davies:1978zz,MasoodASBukhari:2023ljc}.\\
\indent Between two successive evaporations, the black hole reduces in size from an initial horizon radius $r_i$ to a final radius $r_f$. The corresponding change in free energy is given by $\Delta F = F_{\rm f} - F_{\rm i}$. Quantum work is generally defined through the expression \cite{Jarzynski:1996oqb}
\begin{eqnarray}\label{QW1}
    \left\langle \exp{\left(-\frac{W}{T_{\rm BH}}\right)}\right\rangle=\exp{\left(-\frac{\Delta F}{T_{\rm BH}}\right)}.
\end{eqnarray}
This expression essentially relates the average work $W$ done on the black hole system under non-equilibrium conditions to the equilibrium free energy $F$. Furthermore, the Jensen inequality $\langle e^{x} \rangle \geq e^{\langle x \rangle}$ guarantees that $\langle W \rangle - \Delta F \geq 0$, which reflects the second law of thermodynamics\cite{Iso:2011gb}. \textcolor{black}{In broader frameworks, quantum work has been proposed as a tool to explore information preservation in Hawking effect \cite{Pourhassan:2021mhb}}.\\
\indent Now, assuming that the difference in free energy $\Delta F$ in Eq. (\ref{QW1}) depends on the horizon radius $r_{+}$, we can graphically plot the quantum work in Fig. \ref{QW}(b) as a function of $r_{+}$. It is important to note that this provides an effective description of the quantum work, as we do not incorporate full quantum corrections to the black hole geometry. From Fig. \ref{QW}(b), we observe that quantum work is negligible for larger geometries ($r_{+}$), which is expected since fluctuating microstates only become significant in the quantum regime. This can be alternatively understood by examining the expression in Eq. (\ref{QW1}) for larger $r_{+}$, where changes in free energy $\Delta F$ are large, while the Hawking temperature is very small, rendering quantum work negligible. For microscopic scales, quantum work exhibits complex behavior. The first peak, as $r_{+}$ decreases, likely arises due to high values of $T_{\rm BH}$ in that regime, while $\Delta F$ changes slowly in comparison to $T_{\rm BH}$. It vanishes for a certain radius $r_{+}$ when the black hole stops evaporating, corresponding to $T_{\rm BH} = 0$. Beyond this regime, finite values of quantum work persist, which are primarily due to $\eta$, and in a certain sense, reflect the remnant contributions from spacetime fluctuations.

\section{\label{sec:Ruppgeometry} Black hole phase structure via information geometry}
\subsection{Basic tenets of Ruppeiner geometry}

Information or thermodynamic geometry, or in short, \textit{geometrothermodynamics}, \textcolor{black}{provides tools} for understanding phase transitions and the stability of systems undergoing fluctuations around thermal equilibrium. In this approach, a parameter space, akin to Riemannian geometry in gravitational systems, is defined, spanning some extensive quantities of the system, which later aids in defining a Ricci-like curvature \cite{Ruppeiner:1995zz}.  
The initial impetus came from Weinhold \cite{doi:10.1063/1.431689, doi:10.1063/1.431635}, who defined the metric by taking the Hessian of the internal energy with respect to the other extensive variables of the system. This was followed by a rigorous approach by Ruppeiner \cite{Ruppeiner:1979bcp,Ruppeiner:1981znl,Ruppeiner:2013yca}, who employed entropy instead of internal energy. While these methods have been primarily developed and rigorously applied in various well-known fluctuating systems such as quantum liquids, magnetic systems, Ising models, and so on \cite{Ruppeiner:1979bcp,Ruppeiner:1981znl,Janyszek:1989zz,Ruppeiner:2013yca}, \textcolor{black}{their applicability however spans more exotic systems, such as  black holes \cite{Ruppeiner:2013yca}}.
It is anticipated that such a construction may potentially offer insights into the microscopic structure of black hole thermodynamics, which is typically absent in the conventional Bekenstein-Hawking formalism.\\
\indent Denoting the internal energy by $M$, Weinhold metric has the form $g_{\mu\nu}^{W}=\partial_{\mu}\partial_{\nu}M(S,N^i),$
where $S$ is the entropy, while $N^{i}$ are all other extensive quantities indexed by $i$. These quantities may include volume, internal energy etc. Each combination of $\mu,\nu=0,1,2,\dots$ represents one of these quantities. With this, Weinhold line element is given by $\mathrm{d}s_{W}^2=g_{\mu\nu}^{W}\mathrm{d}x^{\mu}\mathrm{d}x^{\nu}.$
Likewise, we have Ruppeiner metric given by $g_{\mu\nu}=-\frac{\partial^2 S }{\partial x^\mu \partial x^\nu}.$ Several other information geometric methods have been developed recently, such as Quevedo \cite{Quevedo:2006xk, Quevedo:2008xn} and the Hendi-Panahiyan-Eslam-Panah-Momennia (HPEM) metric \cite{Hendi:2015rja}, which have also proven useful for exploring black hole thermodynamics. However, our focus here is on employing the Ruppeiner formalism, for which we will now provide a detailed derivation \textcolor{black}{for context and clarity.}\\
\indent Let's start with the standard Boltzmann entropy formula $S=k_{\rm B}\ln\Omega$, with $\Omega$ being the microstate count $\Omega=\exp{\left(S/k_{\rm B}\right)}.$
Next, we consider a parameter space comprising $x^0$ and $x^1$ that define the black hole. As fluctuations occur in the system, we can estimate the probability of finding the black hole system within the intervals $x^0 + \mathrm{d}x^0$ and $x^1 + \mathrm{d}x^1$ as $P(x^0,x^1)\mathrm{d}x^0\mathrm{d}x^1= \Lambda \Omega(x^0,x^1)\mathrm{d}x^0\mathrm{d}x^1$,
where $\Lambda$ is a normalization constant. Using the expression for $\Omega$ given above, we may write $ P(x^0,x^1)\propto \exp{\left(S/k_{\rm B}\right)}$ and 
$S(x^0,x^1)=S_{\rm BH}(x^0,x^1)+S_{\rm E}(x^0,x^1),$
with $S_{\rm BH}$ and $S_{\rm E}$ respectively being the black hole and environment entropies. If there is a small fluctuation in the equilibrium entropy around the point $x_{0}^\mu$ (with $\mu,\nu=0,1$), one can Taylor expand the total entropy around $x_{0}^\mu$ as follows: 
\begin{multline}\nonumber
 S=S_{0}+\frac{\partial S_{\rm BH}}{\partial x^\mu}\bigg|_{x^\mu=x_{0}^\mu}\Delta x_{\rm bh}^\mu+\frac{\partial S_{\rm E}}{\partial x^\mu}\bigg|_{x^\mu=x_{0}^\mu}\Delta x_{\rm E}^\mu \\
 +\frac{1}{2}\frac{\partial ^2 S_{\rm BH}}{\partial x^\mu\partial x^\nu}\bigg|_{x^\mu=x_{0}^\mu}\Delta x_{\rm bh}^{\mu}\Delta x_{\rm bh}^{\nu}\\
 +\frac{1}{2}\frac{\partial ^2 S_{\rm E}}{\partial x^\mu\partial x^\nu}\bigg|_{x^\mu=x_{0}^\mu}\Delta x_{\rm E}^{\mu}\Delta x_{\rm E}^{\nu}+\cdots,
\end{multline}
where $S_{0}$ is the equilibrium entropy at $x_{0}^\mu$.  Now, if we assume a closed system where the extensive parameters of the black hole ($x_{\rm bh}^\mu$) and the environment ($x_{\rm E}^\mu$) have a conservative additive nature, such that $x_{\rm bh}^\mu + x_{\rm E}^\mu = x_{\rm total}^\mu = \text{constant}$, then we can write:
\begin{eqnarray}
 \frac{\partial S_{\rm BH}}{\partial x^\mu}\bigg|_{x^\mu=x_{0}^\mu}\Delta x_{\rm bh}^\mu=-\frac{\partial S_{\rm E}}{\partial x^\mu}\bigg|_{x^\mu=x_{0}^\mu}\Delta x_{\rm E}^\mu.
\end{eqnarray}
We therefore have
\begin{multline}\label{BHE}
\Delta S =\frac{1}{2}\frac{\partial^2 S_{\rm BH}}{\partial x^\mu \partial x^\nu}\bigg|_{x^\mu=x_{0}^\mu}\Delta x_{\rm bh}^{\mu}\Delta x_{\rm bh}^{\nu}\\
+\frac{1}{2}\frac{\partial^2 S_{\rm E}}{\partial x^\mu \partial x^\nu}\bigg|_{x^\mu=x_{0}^\mu}\Delta x_{\rm E}^{\mu}\Delta x_{\rm E}^{\nu}.
\end{multline}
As the entropy of the environment is almost equal to the total entropy, i.e., $S_{\rm E} \sim S$, the corresponding fluctuations in $S_{\rm E}$ are negligible. Consequently, we are left with only the black hole system, such that
$
 P(x^0,x^1)\propto \exp{\left(-\frac{1}{2}\Delta l^2\right)},
$
with $\Delta l^2$ given by $\Delta l^2=-\frac{1}{k_{\rm B}}g_{\mu\nu}\Delta x^{\mu}\Delta x^{\nu}.$
Setting  $k_{\rm B}=1$, one has $\Delta l^2=g_{\mu\nu}\Delta x^{\mu}\Delta x^{\nu}$,
where $g_{\mu\nu}=-\frac{\partial^2 }{\partial x^\mu \partial x^\nu}S_{\rm BH}.$
Given that probability is a dimensionless scalar, $\Delta l^2$, as given up, is a dimensionless, positive definite, invariant quantity. This line element mimics the line element in black holes and is usually considered a quantifying measure of the thermodynamic length between two fluctuating black hole microstates. Quoting Ruppeiner \cite{Ruppeiner:2013yca}: {``Thermodynamic states are further apart if the fluctuation probability is less.''} This principle resonates with Le Chatelier’s principle, which ensures the local stability of thermodynamic systems. Dropping the subscript $\rm BH$, we write $g_{\mu\nu}=-\frac{\partial^2 }{\partial x^\mu \partial x^\nu}S$, as the metric of Ruppeiner geometry. Based on this metric, we can now compute the associated curvature scalar in the same fashion as one usually does in Riemannian geometry. Given the Christoffel connections are 
$\Gamma_{\mu\nu}^\sigma=\frac{1}{2}g^{\sigma\rho}\left(\partial_{\nu}g_{\rho\mu}+\partial_{\mu}g_{\rho\nu}-\partial_{\rho}g_{\mu\nu}\right)$,
along with the Riemann tensor
$R_{\rho\mu\nu}^\sigma=\partial_{\nu}\Gamma_{\rho\mu}^\sigma-\partial_{\mu}
\Gamma_{\rho\nu}^{\sigma}+\Gamma_{\rho\mu}^{\delta}\Gamma_{\delta\nu}^{\sigma}-\Gamma_{\rho\nu}^{\delta}\Gamma_{\delta\mu}^{\sigma}$,
one can define  Ricci tensor and scalar as 
$R_{\mu\nu}=R_{\mu\sigma\nu}^{\sigma}$ and $R=g^{\mu\nu}R_{\mu\nu}$. 
Here, the Ricci curvature is \cite{Carroll:2004st}
\begin{align}\label{Rcurvature}
\begin{aligned}
R&= -\frac{1}{\sqrt{g}}\left[ \frac{\partial}{\partial x^0} \left(\frac{g_{01}}{g_{00}\sqrt{g}}\frac{\partial g_{00}}{\partial x^1}-\frac{1}{\sqrt{g}}\frac{\partial g_{11}}{\partial x^0}\right) \right.  \\[6pt]
& \left. +\frac{\partial}{\partial x^1}\left(\frac{2}{\sqrt{g}}\frac{\partial g_{01}}{\partial x^0}-\frac{1}{\sqrt{g}}\frac{\partial g_{00}}{\partial x^1}-\frac{g_{01}}{g_{00}\sqrt{g}}\frac{\partial g_{00}}{\partial x^0}\right)\right],
\end{aligned}
\end{align}
with $ g := \det{g_{\mu\nu}} = g_{00} g_{11} - g_{01}^2 $. 

\subsection{Computing the Ruppeiner thermodynamic curvature $R_{C}$}
Employing the same technique for the Ruppeiner metric 
$g_{\mu\nu}=-\partial_{\mu}\partial_{\nu}S(M,N^i)$, with  a $2$-dimensional state space of non-diagonal $g_{\mu\nu}$, the line element is given by
\begin{eqnarray}
 \mathrm{d}s_{R}^2=g_{MM}\mathrm{d}M^2+2g_{MQ}\mathrm{d}M\mathrm{d}Q+g_{QQ}\mathrm{d}Q^2,
\end{eqnarray}
with the metric $g_{\mu\nu}$ specified as $g_{00}=g_{MM},g_{01}=g_{MQ},g_{10}=g_{QM},g_{11}=g_{QQ},$
where we have used $M$ and $Q$ as the extensive variables as they are the most natural choice for our charged black hole.
These components of $g_{\mu\nu}$ can be computed by expressing the metric in terms of derivatives of the entropy with respect to the extensive variables as follows: 
\begin{eqnarray}
 g_{MM}=-\frac{\partial}{\partial M}\left(\frac{\partial S}{\partial M}\right),\ g_{MQ}=-\frac{\partial}{\partial M}\left(\frac{\partial S}{\partial Q}\right),\\
 g_{QM}=-\frac{\partial}{\partial Q}\left(\frac{\partial S}{\partial M}\right), \ g_{QQ}=-\frac{\partial}{\partial Q}\left(\frac{\partial S}{\partial Q}\right),
\end{eqnarray}
which can be explicitly evaluated, as shown in the Appendix.
We are now in a position to compute the thermodynamic curvature 
   \begin{align}\nonumber
R_{C}&= -\frac{1}{\sqrt{g}}\bigg[ \frac{\partial}{\partial M} \left(\frac{g_{MQ}}{g_{MM}\sqrt{g}}\frac{\partial g_{MM}}{\partial Q}-\frac{1}{\sqrt{g}}\frac{\partial g_{QQ}}{\partial M}\right)  \\ 
\label{RC} &+\frac{\partial}{\partial Q}\left(\frac{2}{\sqrt{g}}\frac{\partial g_{MQ}}{\partial M}-\frac{1}{\sqrt{g}}\frac{\partial g_{MM}}{\partial Q}-\frac{g_{MQ}}{g_{MM}\sqrt{g}}\frac{\partial g_{MM}}{\partial M}\right)\bigg],
\end{align}
where $g=\det{g_{\mu\nu}}=g_{MM}g_{QQ}-g_{MQ}^2$. \\
\indent \textcolor{black}{In this framework, the following interpretation is usually ascribed to $R_{C}$}: a zero curvature indicates that the system is non-interacting and in an ideal gas-like configuration, providing no additional information about black hole micromolecules.  \textcolor{black}{A positive $R$ is associated with repulsive effective interactions, while negative $R$ indicates attractive interactions.} Divergences in $R$ correspond to phase transitions \cite{Ruppeiner:2013yca}.
From this interpretation of $R$, one might wonder if $R$ would be very large for a black hole due to its infinite density. This point is emphasized in Ref. \cite{Ruppeiner:2013yca}, where it is suggested that the gravitational degrees of freedom in a black hole system might possess some non-statistical description because all the gravitating material has been compressed into the central singularity. Thus, thermodynamic curvature represents some kind of non-gravitational interactions among the black hole constituents at its surface, arising effectively from the underlying gravitational degrees of freedom.

\begin{figure*}[t]
\centering
\includegraphics[width=\linewidth, height=11cm]{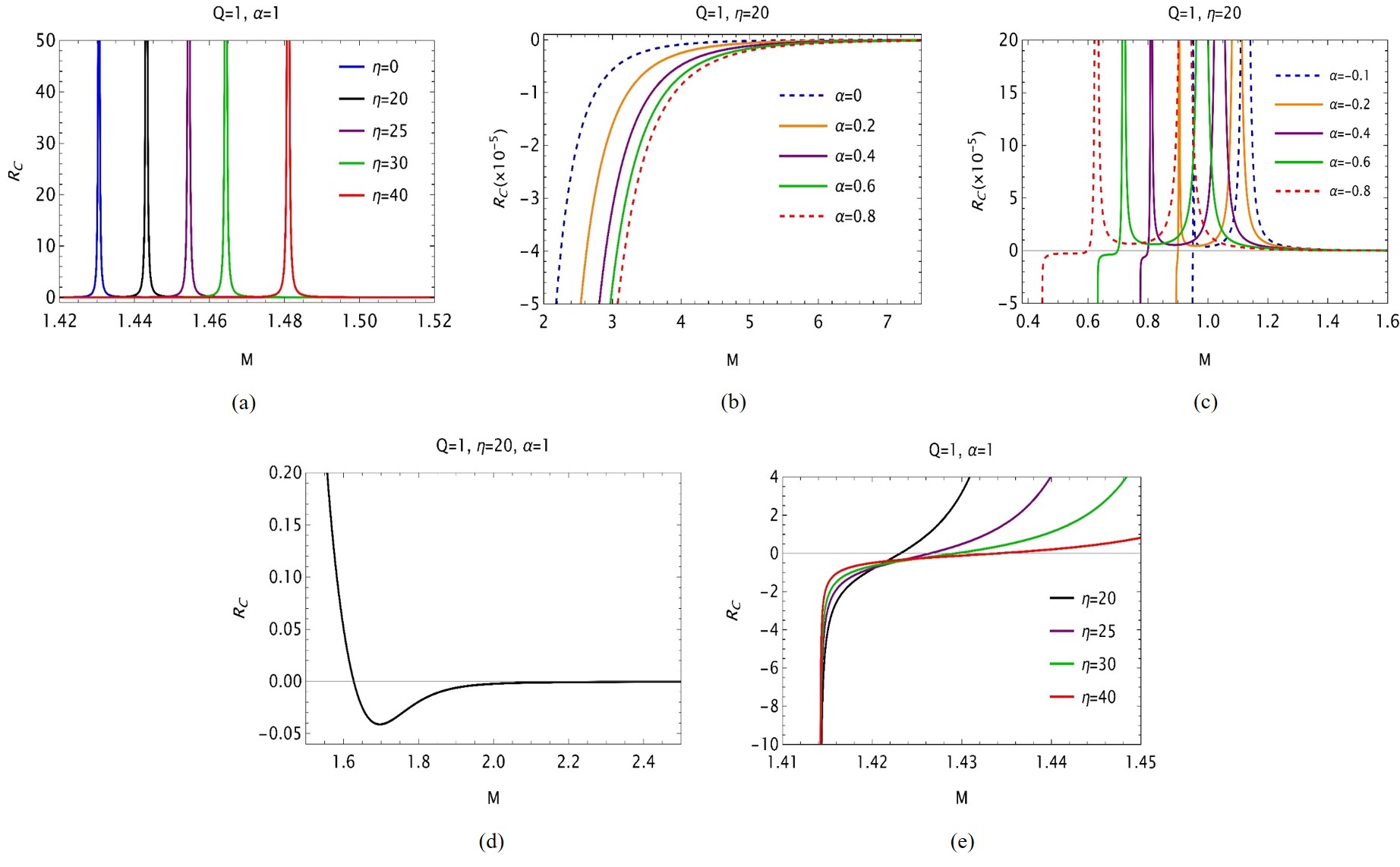}
\caption{\textcolor{black}{Thermodynamic curvature $ R_C $ versus black hole size: impact of (a) quantum corrections $ \eta $, (b) positive GB coupling parameter $ \alpha $, and (c) negative GB coupling parameter $ \alpha $; (d) zoomed-in view of plot (a) with $ \eta = 20 $, and (e) zoomed-in view near the extremal limit $ M = \sqrt{Q^2 + \alpha} $.}}
\label{RCplot}
\end{figure*}

\subsection{\textcolor{black}{Insights from the curvature scalar}}

\textcolor{black}{The expression for the thermodynamic curvature $R_{C}$ obtained from Eq.~(\ref{RC}) is lengthy, therefore we refrain from writing it down here; we instead summarize the features extracted from Fig.~\ref{RCplot} for representative values of $(\eta,\alpha)$. A robust property across all panels is that $R_{C}\to 0$ for large black hole sizes (large $M$), essentially independent of $(\eta,\alpha)$. In the Ruppeiner interpretation this corresponds to a flat, ideal-gas-like (non-interacting) thermodynamic sector. Such Ruppeiner flatness is known for charged black holes in Einstein gravity across all spacetime dimensions~\cite{Aman:2003ug,Aman:2005xk,MasoodASBukhari:2023ljc}, and its persistence here provides a direct GR baseline: deviations appear only in the small-horizon/near-extremal regime where $(\alpha,\eta)$ become relevant.}\\
\indent \textcolor{black}{As $M$ decreases from the macroscopic regime, $R_{C}$ departs from zero and develops curvature singularities that partition the state space into distinct regions. For fixed $Q$ and $\alpha$, Fig.~\ref{RCplot}(a) shows two divergences: a positive divergence at a larger value of $M$, and a negative divergence at the extremal point where $M=\sqrt{Q^{2}+\alpha}$. Between these poles, $R_{C}$ becomes negative [cf.\ Fig.~\ref{RCplot}(d)], indicating an effective attractive-interaction signature in the usual Ruppeiner sign convention before a repulsive ($R_{C}>0$) sector is encountered around the positive pole.}\\
\indent \textcolor{black}{The positive-$R_{C}$ region and its divergence signal a critical point (phase-transition-like behavior in the information-geometric diagnostic). As the strength of the entropy correction $\eta$ increases, the positive divergence shifts to larger $M$ [Fig.~\ref{RCplot}(a)]. By contrast, the negative divergence remains tied to the endpoint $M=\sqrt{Q^{2}+\alpha}$ for each curve [Fig.~\ref{RCplot}(e)], reflecting that the terminal point is fixed by the background extremality condition. Increasing $\alpha$ at fixed $(Q,\eta)$ shifts this endpoint to larger $M$ [Fig.~\ref{RCplot}(b)], moving the termination of the fixed-$Q$ evaporation picture in $M$--space relative to the GR/RN baseline ($M=|Q|$). For $\alpha<0$, $\sqrt{Q^{2}+\alpha}$ decreases and the negative divergence occurs at smaller $M$ [Fig.~\ref{RCplot}(c)]; in addition, two positive divergences appear in this range, indicating additional critical points.}\\
\indent \textcolor{black}{A further important observation is that the near-extremal region remains in the negative-$R_{C}$ sector for all cases shown (including $\alpha<0$), while $R_{C}\to -\infty$ at the extremal boundary. Moreover, the   divergence in $R_{C}$ in the extremal limit coincides with the vanishing of $C_{Q}$ at extremality [Fig.~\ref{CQ}], providing a nontrivial consistency check between the information-geometric and response-function diagnostics. We therefore interpret the Ruppeiner curvature here as an effective probe of correlation/interaction signatures encoded by the quantum-corrected entropy, while the canonical-ensemble stability information is captured by $C_{Q}$.}\\
\indent \textcolor{black}{The extremal endpoint also corresponds to $T_{BH}\to 0$ [Fig.~\ref{Tplot}] and hence to a remnant-like termination of the evaporation picture in our fixed-$Q$ framework. Since $C_{Q}$ and $R_{C}$ become unphysical beyond extremality in this setup, we do not extend the analysis past the endpoint. Nevertheless, it is suggestive to compare with ordinary low-temperature thermodynamics: as $T_{BH}\to 0$, interactions can freeze out, leaving behind quantum-statistical interactions as in ideal Fermi or Bose gases \cite{Ruppeiner:2023wkq}. In our case, the divergence $R_{C}\to -\infty$ at $T_{B}=0$ may therefore be read heuristically as an extreme attractive-correlation signature compatible with a frozen, remnant-like configuration.}\\ 
\indent \textcolor{black}{A microscopic explanation of these features remains challenging. Black hole thermodynamics is itself subtle at the microscopic level \cite{Davies:1978zz, Page:2004xp}, and the simultaneous presence of higher-curvature corrections (through $\alpha$) and short-distance entropy deformations (through $\eta$) complicates a direct microscopic interpretation. We may, however, still infer certain qualitative  features from the present ansatz. From Eq.~\ref{root} (or Eq.~\ref{rpm1}), the horizon scale is primarily controlled by $M$, so the influence of $\alpha$ becomes pronounced only at small radii (remember $\alpha$ is tightly constrained phenomenologically \cite{Fernandes:2022zrq}). In this sense, $\alpha$ behaves in a $Q$-like manner in shrinking the horizon scale relative to the GR limit, thereby shifting the near-extremal regime where the thermodynamic-geometry signatures become nontrivial. The macroscopic Ruppeiner flatness may be viewed as reflecting an effective cancellation of attractive and repulsive interaction signatures, while at quantum scales one sign can dominate; in our plots the near-extremal sector is consistently dominated by the attractive ($R_{C}<0$) signature.}\\
\indent \textcolor{black}{Finally, we emphasize that standard Ruppeiner geometry is rooted in equilibrium thermodynamic fluctuations, whereas at quantum scales non-equilibrium ingredients may become relevant \cite{Pourhassan:2021mhb,Pourhassan:2022irk,Iso:2011gb,Pourhassan:2020bzu}. Here we incorporate quantum effects through an entropy deformation without modifying the spacetime geometry, so the curvature analysis should be interpreted as an effective thermodynamic diagnostic based on a quantum-corrected entropy rather than a fully microscopic derivation. For related discussions of repulsive/attractive microstructure interpretations using Ruppeiner geometry in GB gravity, see, e.g., refs. \cite{Wei:2020poh,Wei:2019uqg}.}

\subsection{\textcolor{black}{Theoretical and observational prospects}}
\label{sec:implications}
\indent \textcolor{black}{In the early-Universe context, near-extremal endpoints with $T_{BH}\to 0$ are relevant for primordial black holes (PBHs)~\cite{Carr:1974nx}.  In semiclassical GR, sufficiently light PBHs would evaporate completely, whereas an evaporation endpoint at finite mass would instead leave long-lived relics~\cite{Carr:2020gox}. Such PBH relic/remnant scenarios have been discussed as potential contributors to (or constituents of) dark matter~\cite{Dalianis:2019asr,Carr:2021bzv} and are constrained by a broad set of cosmological and astrophysical bounds across PBH mass ranges~\cite{Carr:2020gox}. In our effective setup, the 4D-EGB background shifts the extremality bound from the GR/RN value to $M=\sqrt{Q^{2}+\alpha}$, implying a modified terminal mass scale (and thus a modified remnant endpoint) in a fixed-$Q$ evaporation picture. This suggests that, if a similar shift persists in a fully dynamical treatment (including charge loss and backreaction), higher-curvature effects could alter the parameter space in which PBH remnants are produced and survive~\cite{Sasaki:2025zao}. A natural next step is therefore to embed the present thermodynamic diagnostics into an early-Universe PBH evolution model and juxtapose the shifted-endpoint scenario with existing PBH abundance constraints~\cite{Carr:2020gox,Carr:2021bzv}.\\
\indent Beyond these cosmological implications, $\alpha$ also interfaces with strong-field tests of gravity in the compact-object sector. In particular, it is useful to treat $\alpha$ as an effective strong-field parameter whose physical interpretation (and hence constraining process) is tied up  with specific four-dimensional completion/regularization adopted for the underlying 4D-EGB construction~\cite{Fernandes:2022zrq}. A concrete motivation for focusing on strong-field probes is that, in 4D-EGB black hole spacetimes, quasinormal-mode spectra and photon-sphere/shadow observables acquire explicit $\alpha$-dependence, while stability considerations can already impose nontrivial restrictions on the admissible range of $\alpha$ (see~ref. \cite{Konoplya:2020bxa} and references therein). These are precisely the sectors that connect most directly to current and near-future observational programs, including gravitational-wave ringdown measurements~\cite{LIGOScientific:2016aoc} and horizon-scale imaging by the Event Horizon Telescope~\cite{EventHorizonTelescope:2019dse}. In parallel, astrophysical consistency studies performed in scalar--tensor realizations motivated by higher-dimensional GB gravity indicate that demanding viable compact objects across masses and radii can yield stringent, yet model-dependent, restrictions on the effective coupling~\cite{Charmousis:2021npl}. Within this landscape, our thermodynamic analysis plays a complementary role: rather than attempting to extract bounds directly, it highlights the region of state space where thermodynamic response functions and information-geometric signatures become maximally sensitive to $\alpha$, thereby indicating where a fully dynamical strong-field confrontation with data (including backreaction, charge loss, and/or quantum-corrected geometry) would be most informative.}

 \setlength{\parskip}{0.3cm}
    \setlength{\parindent}{1em}

\section{\label{sec:summary} Conclusion and Outlook}
\setlength{\parskip}{0.1cm}
\textcolor{black}{In this work, we investigated the thermodynamic phase structure of a charged 4D-EGB black hole in the GR branch, incorporating quantum-gravity-motivated non-perturbative (exponential) corrections to the entropy while keeping the spacetime geometry, and hence the Hawking temperature defined by surface gravity, classical. Within this effective setup (fixed $Q$ canonical picture), we combined standard thermodynamic diagnostics (heat capacity and Helmholtz free energy) with two complementary probes: (i) an effective quantum-work functional derived from the free-energy landscape via Jarzynski equality, and (ii) information-geometric diagnostics based on the Ruppeiner scalar curvature.\\
\indent Our results indicate that in the macroscopic regime, the response functions approach the corresponding general-relativistic behavior, consistent with the expectation that both GB and entropy-deformation effects ($\eta$) are suppressed at large horizon scales. In contrast, in the small-horizon/near-extremal regime---where the exponential entropy deformation becomes relevant---the interplay between the GB coupling ($\alpha$) and  $\eta$ can generate additional thermodynamic instabilities and curvature singularities. In particular, the endpoint of the fixed-$Q$ evolution is controlled by the extremal limit $M=\sqrt{Q^{2}+\alpha}$, which is shifted relative to the RN baseline ($\alpha=0$) and thereby modifies the inferred terminal mass scale and the associated stability diagnostics in a parameter-dependent manner. The information-geometric analysis provides an internally consistent cross-check: divergences of the Ruppeiner curvature track the loci of thermodynamic singularities, while the sign of the curvature offers a qualitative diagnostic of effective correlations in the thermodynamic state space. At the same time, since quantum effects are implemented here through an entropy deformation on a classical background, any microscopic interpretation should be regarded as model-dependent. \\
\indent The $\alpha$-shifted extremality bound $M=\sqrt{Q^{2}+\alpha}$ implies a modified near-extremal endpoint in our fixed-$Q$ effective picture. 
If this feature persists in a fully dynamical PBH evolution (including charge loss and backreaction), it could affect PBH relic formation and should be tested against existing abundance constraints. 
Since $\alpha$ also enters strong-field observables (e.g., quasinormal modes and photon-sphere/shadow properties), the near-extremal regime provides a promising target for future constraints.
\\
\indent Several extensions are natural. It would be of interest to generalize the analysis to rotating solutions and to include a negative cosmological constant in an extended thermodynamic framework, where additional phase structure and Hawking-Page-type phenomena may arise \cite{Kubiznak:2016qmn,Hawking:1982dh}. More fundamentally, a fully dynamical assessment would require incorporating quantum-corrected geometry and backreaction effects to test the robustness of the near-extremal behavior and remnants reported in this study.}

\section*{Data Availability Statement}
No new data were generated in this study; all results follow from the analytical calculations presented in the manuscript.

 \section*{\label{appendix}Appendix}
 The metric elements are obtained as follows:
\begin{widetext}
\begin{align*}
g_{MM}=&-\frac{1}{\left(
M^2-Q^2-\alpha\right)^{3/2}}\bigg[ 2 e^{-\left(\sqrt{ M^2-Q^2-\alpha}+M\right)^2} \\
& \times \left(\sqrt{ M^2-Q^2-\alpha}+M\right)^2
 \bigg\{e^{\left(\sqrt{ M^2-Q^2-\alpha}+M\right)^2}
 \left(2 \sqrt{ M^2-Q^2-\alpha}-M\right)\\
  &+\eta  \left[4 M^3+4 M^2 \sqrt{ M^2-Q^2-\alpha}-2 \left(\alpha +Q^2+1\right) \sqrt{ M^2-Q^2-\alpha}+M \left(-4 \alpha -4 Q^2+1\right)\right]\bigg\}\bigg],\\
g_{MQ}=&\frac{2 Q}{\left(M^2-Q^2-\alpha\right)^{3/2}}
\bigg[-\alpha +\eta  e^{-\left(\sqrt{ M^2-Q^2-\alpha}+M\right)^2} \bigg\{\alpha +2 \Big[4 M^4-5 M^2 \left(\alpha +Q^2\right)\\
&-3 M \left(\alpha +Q^2\right) \sqrt{ M^2-Q^2-\alpha} +4 M^3 \sqrt{ M^2-Q^2-\alpha}+\left(\alpha +Q^2\right)^2\Big]+Q^2\bigg\}-Q^2\bigg],\\
 g_{QM}=&\frac{2 Q}{\left(M^2-Q^2-\alpha\right)^{3/2}} \bigg[-\alpha +\eta  e^{-\left(\sqrt{ M^2-Q^2-\alpha}+M\right)^2} \bigg\{\alpha +2 \Big[4 M^4-5 M^2 \left(\alpha +Q^2\right)\\
 &-3 M \left(\alpha +Q^2\right) \sqrt{ M^2-Q^2-\alpha}
  +4 M^3 \sqrt{ M^2-Q^2-\alpha}+\left(\alpha +Q^2\right)^2\Big]+Q^2\bigg\}-Q^2\bigg],\\
g_{QQ}=&\frac{2}{\left( M^2-Q^2-\alpha\right)^{3/2}} \bigg[M^3+M^2 \sqrt{ M^2-Q^2-\alpha}-\left(\alpha +Q^2\right) \sqrt{ M^2-Q^2-\alpha}+\eta  e^{-\left(\sqrt{ M^2-Q^2-\alpha}+M\right)^2}\\
&\Big\{-M^3 \left(4 Q^2+1\right)+\left[ \left(2 Q^2+1\right) \left(\alpha +Q^2\right)-M^2 \left(4 Q^2+1\right)\right] \sqrt{M^2-Q^2-\alpha}+M \left[\alpha +4 Q^2 \left(\alpha
   +Q^2\right)\right]\Big\}-\alpha  M\bigg],
\end{align*}
along with the determinant 
\begin{align*}
    g&=-\frac{4 e^{-2 \left(\sqrt{ M^2-Q^2-\alpha}+M\right)^2} }{\left(M^2-Q^2-\alpha\right)^{5/2}} \Bigg[\Big(e^{\left(\sqrt{ M^2-Q^2-\alpha}+M\right)^2}-\eta \Big)e^{\left(\sqrt{ M^2-Q^2-\alpha}+M\right)^2} \bigg\{4 M^5-M^3 \left(9 \alpha +7 Q^2\right)-M^2 \left(7 \alpha +5 Q^2\right) \\
    & \times \sqrt{ M^2-Q^2-\alpha}+\left(\alpha +Q^2\right) \left(2 \alpha +Q^2\right) \sqrt{ M^2-Q^2-\alpha}+4M^4 \sqrt{ M^2-Q^2-\alpha}+M \left(\alpha +Q^2\right) \left(5 \alpha +3 Q^2\right)\bigg\}\\
    &+\eta  \bigg\{32 M^7-4 M^5 \left(18 \alpha +18 Q^2+1\right)+M^3 \left[\alpha  (50 \alpha +9)+50 Q^4+(100 \alpha +7) Q^2\right]+M^2 \left[\alpha  (26 \alpha +7)+26 Q^4+(52 \alpha +5) Q^2\right]\\
    & \times \sqrt{ M^2-Q^2-\alpha}-\left(\alpha +Q^2\right) \left[2 \alpha 
   (\alpha +1)+2 Q^4+(4 \alpha +1) Q^2\right] \sqrt{ M^2-Q^2-\alpha}+32 M^6 \sqrt{ M^2-Q^2-\alpha}-4 M^4\\
   & \times \left(14 \alpha +14 Q^2+1\right) \sqrt{ M^2-Q^2-\alpha}-M \left(\alpha +Q^2\right) \left[5 \alpha  (2 \alpha +1)+10 Q^4+(20 \alpha +3) Q^2\right]\bigg\}\Bigg].
\end{align*}
\end{widetext}

\bibliographystyle{apsrev4-1}
\bibliography{masood.bib}
\end{document}